\documentclass[%
 aip,
 amsmath,amssymb,
reprint,%
]{revtex4-1}

\usepackage{graphicx}
\usepackage{dcolumn}
\usepackage{bm}
\usepackage{siunitx,chemformula}
\usepackage{textgreek}

\usepackage[utf8]{inputenc}
\usepackage[T1]{fontenc}
\usepackage{mathptmx}
\usepackage{etoolbox}
\usepackage{xcolor}
\usepackage{booktabs}
\makeatletter
\def\@email#1#2{%
 \endgroup
 \patchcmd{\titleblock@produce}
  {\frontmatter@RRAPformat}
  {\frontmatter@RRAPformat{\produce@RRAP{*#1\href{mailto:#2}{#2}}}\frontmatter@RRAPformat}
  {}{}
}%
\makeatother

\begin{document}

\preprint{XXX}

\title[3D etching]{Plasma sheath tailoring by a magnetic field for three-dimensional plasma etching}
\author{Elia Jüngling}
\affiliation{ 
Chair Experimental Physics II, Ruhr University Bochum, Bochum, Germany
}%
\author{Sebastian Wilczek}%
\affiliation{ 
Chair of Applied Electrodynamics and Plasma Technology, Ruhr University Bochum, Bochum, Germany
}%

\author{Thomas Mussenbrock}%
\affiliation{ 
Chair of Applied Electrodynamics and Plasma Technology, Ruhr University Bochum, Bochum, Germany
}%

\author{Marc Böke}%
\affiliation{ 
Chair Experimental Physics II, Ruhr University Bochum, Bochum, Germany
}%
\author{Achim von Keudell}%
\affiliation{ 
Chair Experimental Physics II, Ruhr University Bochum, Bochum, Germany
}%

\date{\today}


\begin{abstract}
Three-dimensional (3D) etching of materials by plasmas is an ultimate challenge in microstructuring applications. A method is proposed to reach a controllable 3D structure by using masks in front of the surface in a plasma etch reactor in combination with local magnetic fields to steer the incident ions in the plasma sheath region towards the surface to reach 3D directionality during etching and deposition. This effect can be controlled by modifying the magnetic field and/or plasma properties to adjust the relationship between sheath thickness and mask feature size. Since the guiding length scale is the plasma sheath thickness, which for typical plasma densities is at least 10s of microns or larger, controlled directional etching and deposition target the field of microstructuring, e.g. of solids for sensors, optics, or microfluidics. In this proof-of-concept study, it is shown that $\vec{E}\times\vec{B}$ drifts tailor the local sheath expansion, thereby controlling the plasma density distribution and the transport when the plasma penetrates the mask during an RF cycle. This modified local plasma creates a 3D etch profile. This is shown experimentally as well as using 2d3v Particle-In-Cell/Monte Carlo collisions simulation.

\end{abstract}

\maketitle


Plasma processing of materials is very advanced in microelectronics to create 2D structures with very high precision by anisotropic etching in silicon or 2D materials\cite{sekine2002dielectric,radjenovic2014implementation,kruger2019voltage}. This is complemented by various deposition techniques ranging from standard plasma-enhanced chemical vapor deposition (CVD)\cite{orfert1999plasma} to atomic layer deposition (ALD) or plasma-ALD\cite{lim2004method} to generate various oxide or metal layers\cite{Arts.2022}. The directed ion flux ensures the directionality of the etch process in plasma-based microstructuring due to the process of chemical sputtering, where the impact of ions and reactive species simultaneously ensures a high etch rate. This process is e.g. exploited in trench etching, where the bottom of a trench is etched very efficiently compared to the side walls. This effect is also amplified by side wall passivation due to the polymerization of the reactive etching gas. This allows for creating very small structures down to 7\,nm for a gate oxide width or trenches with extremely high aspect ratios in the range of 50 to 100 (depth vs. width). Plasma processing is also well-advanced to create microstructures with feature sizes on the micrometer scale to produce micro-electromechanical systems (MEMS) devices, preferably in silicon for sensors or microfluidic devices. 3D printing technologies are employed at even larger scales using mainly polymethyl methacrylate (PMMA) or polyether ether ketone (PEEK) as material. However, there is demand for advanced 3D structuring of materials for optical applications such as microlenses\cite{Brunner.2004}, gratings\cite{Amako.2009} or for illumination \cite{Miller.1993}. All these applications require 3D shape control with a small surface roughness in the range of $\lambda$/100, with $\lambda$ being the wavelength in the particular application. Another field for 3D structuring is microfluidic fuel cells, where the turnover of an electrochemical reaction is enhanced by replacing a membrane by a diffusion gradient in a microchannel and designing specific 3D electrode surfaces\cite{zhou.2021}.

Realizing complex structures requires accurate 3D control of plasmas since the sidewalls and the bottom and top of the channel must be processed differently. In the past, several attempts have been explored to reach 3D plasma processing capabilities: (i) \textit{Shadowing effects in glancing angle deposition} - the shadowing of the growing structures during glancing angle deposition of metals may lead to micro-pillars on top of a surface\cite{Hawkeye.2007}; (ii) \textit{Charging of 3D structures in the plasma sheath to steer the ions} - Chang et al. \cite{Chang.2020} used structures above the wafer level that charge up and distort the plasma sheath in front of the surface. This steers the ions, and a tilted angle etching and manufacturing of MEMS structures is realized. Yoon et al. \cite{Yoon.2022} used a similar concept to tilt the incident ions; (iii) \textit{Alternating chemistry during trench etching:} - Ni et al. \cite{Ni.2020} reached 3D patterning by alternating the etching of deep holes using the anisotropic etch characteristics of a fluorocarbon etch process, with a subsequent isotropic chemical etch step. This sequence can be repeated several times to reach stacked 3D structures; (iv)  \textit{Affecting trench charging by magnetic fields} - Schaepkens and Oehrlein \cite{Schaepkens.1998} used a weak magnetic field parallel to the wafer surface, which affected the electrons to charge one side of the sidewall. This created an electric field component, which diverted the incident ions in the very same direction inside a trench during plasma etching; (v) \textit{Micromachining for optics} - macroscopic 3D structures can be created by various micromachining techniques such as electrical discharge machining or laser ablation \cite{Pawar.2017,Hof.2017}. These are serial processes operating more on the macroscopic scale; (vi) \textit{Use of photosensitive materials} - the 3D aspect of an etch process can be controlled by using photosensitive materials, where an external illumination changes the material properties so that the etch rates develop locally differently\cite{Weigel.2021}. This requires, however, the choice of particular photosensitive materials.

In this paper, we introduce the following idea for a flexible 3D pattering process: a magnetic field $\vec{B}$ is applied parallel to the surface, leading to $\vec{E} \times \vec{B}$ drifts in combination with the electric fields $\vec{E}$ in the plasma sheaths. A mask is placed in front of the substrate to design the $\vec{E}$ fields and thus allow for a 3D control of the plasma density in front of a substrate by tailoring the plasma sheath with the $\vec{E} \times \vec{B}$ drifts. This local variation of the plasma density, in turn, leads to a well-defined 3D etching pattern.

\begin{figure}
    \centering
    \includegraphics[width=0.6\textwidth]{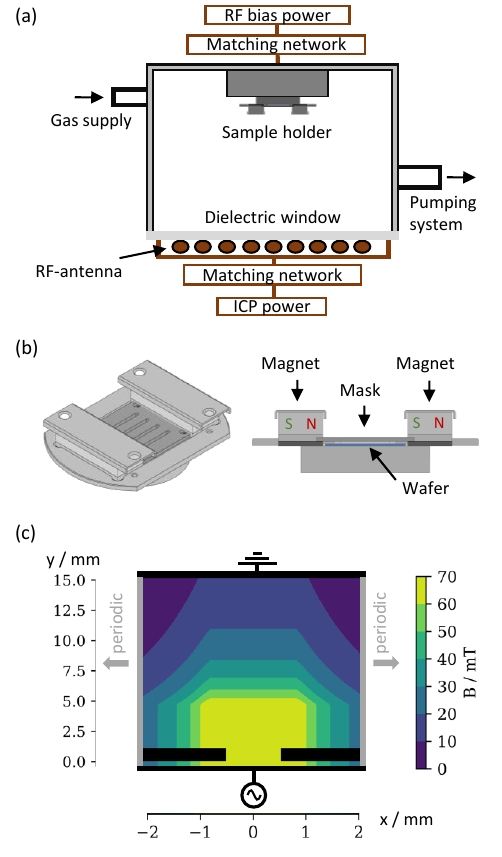}
    \caption{Schematic of the plasma etching setup consisting of an ICP plasma at the bottom and RF-biased substrate holder at the top (a). 3D sketch (left) and cross-section (right) of the sample assembly consisting of a wafer with a mask in front and attached magnets (b). Schematic of the simulation domain, including the magnetic field configuration in the simulation (c). (a) and (b) are not to scale. 
    }
    \label{fig:setup}
\end{figure}

The etching experiments are performed in an inductively coupled plasma (ICP)\cite{chabert2021foundations,chabert2011physics} etching setup (see Fig. \ref{fig:setup}a) using an argon
-CF$_4$ mixture as a processing gas for silicon etching and an argon-oxygen mixture for hydrocarbon film etching, both in a $5:1$ ratio. 
A sample assembly is placed on a substrate holder opposite the ICP dielectric window. At the substrate holder, an external RF bias voltage can be applied. The distance between the RF substrate holder and the ICP dielectric window is 13.5\,cm.  

The sample assembly consists of a silicon wafer (or a silicon wafer coated with a hydrocarbon coating) mounted below a metal grid (thickness 1\,mm) with four linear slits and two adjacent permanent magnets that generate a magnetic field parallel to the grid openings in the order of 70\,mT. The width of the slit is 1mm and the distance in between two slits is 2\,mm. The distance between the grid and the wafer is 0.5\,mm. A schematic illustration is shown in  Fig. \ref{fig:setup}b.

For a typical experiment, the applied ICP power is set to 500\,W for hydrocarbon film etching and 600\,W  for silicon etching, respectively. All experiments are performed at a pressure of 2\,Pa. The RF bias power is automatically adjusted to reach a given DC self-bias. Here, a DC self-bias of -100\,V and -150\,V is used for hydrocarbon film and silicon etching, respectively. Both the ICP power and the RF bias power are operated at a frequency of 13.56\,MHz.

After the sample has been processed for a given time, the assembly is removed from the setup and dismounted to analyze the etch profiles on the wafer below the slits.   

The geometry of the etch profiles is analyzed using profilometry (Veeco 6M, horizontal resolution: 5\,µm, vertical resolution: 2\,nm). Measurements are taken exclusively at the center of the two middle slits because the magnetic field of the sample assembly is uniform only in that location.
Ex-situ XPS measurements were performed to determine the stoichiometry of the samples. The measurements were realized with a PHI5000 x-ray photoelectron spectrometer using Al~K\textalpha~radiation at 1,486.6\,eV. A beam diameter of 200\,µm and a pass energy of \SI{187.85}{\eV} 
were used. The step width was \SI{1}{\eV}, resulting in a resolution of \SI{2.82}{\eV}. 


To analyze the manipulation of the plasma transport in front of the surface by the mask and the magnetic field, a Particle-In-Cell/Monte Carlo collisions (PIC/MCC) simulation is used. The 2d3v PIC/MCC code had been designed following the concept of the 1D code eduPIC\cite{Donko.2021} and has been validated against benchmarks in the literature\cite{Donko.2021, Turner.2013}. An excellent agreement has been found. 
This 2d3v PIC/MCC code simulates a capacitively coupled plasma (CCP)\cite{chabert2011physics,lieberman1994principles,wilczek2020electron} in pure argon to quantify the dynamic asymmetry in the incident ion flux at the wafer surface. The full Lorentz force law $\vec{F} = ±q(\vec{E} + \vec{v} \times \vec{B})$ is considered for both electrons and ions, although the ions are predominantly non-magnetized due to their large mass\cite{Trieschmann.2013}. We use an explicit push scheme based on Boris's approach\cite{Boris.1970}. 
While the simulation considers multiple collision processes in argon using cross sections from literature\cite{Phelps.1999}, any complex plasma chemistry regarding reactive processing gases has been neglected. 
Furthermore, the code includes glancing angle reflection of ions at the inside walls of the mask in front of the wafer. For this, a simple model has been developed using reflection probabilities calculated with SRIM\cite{Ziegler.2010}. 

 The simulation domain is defined as a mask in front of a wafer at a distance of 0.5\,mm with a single 1\,mm slit; the thickness of the mask is 1\,mm. The whole 2D simulation area is 4\,mm x 15\,mm. The magnetic field is adjusted parallel to the wafer surface pointing out of the plane. A schematic illustration of the geometry is shown in Fig.\ref{fig:setup}c. The driving sinusoidal voltage waveform with an amplitude of 100\,V and a frequency of 13.56\,MHz is applied to the lower boundary and the mask elements, while the upper electrode is grounded. Both surfaces and the mask structure, except for the inside walls of the mask where the reflection model is used, fully absorb incident particles. Secondary electron emission at all surfaces is neglected. Periodic boundary conditions are used at the sides of the simulation domain. The computational grid is Cartesian and equidistant with a cell width of 0.1\,mm and a time step of 35\,ps is used. The initial particle configuration consists of $10^5$ electrons and ions with an initial temperature of 3\,eV and 300\,K, respectively, and a particle weight of $8 \cdot 10^6$. Consequently, the conditions for stability and accuracy are fulfilled with these parameters \cite{Donko.2021,Turner.2013}. 
 The background gas is set to a pressure of 2\,Pa. The time- and space-resolved results are acquired after 50\,µs of simulation time when convergence is reached. The ion flux is time-integrated from 50\,µs to 100\,µs simulation time.
 The magnetic field is designed to affect only a small region near the slit (extending only $\pm$ 0.5\,mm from both sides of the slit). This local restriction of the magnetic field is artificial, and therefore only the modeling results in the vicinity of the slit region are reliable. At first, a homogeneous magnetic field in the complete simulation area was tested, which led to excellent confinement of the electrons because the magnetic field parallel to the surface reduced the cross-magnetic field transport. Still, the code did not reach a stationary solution. Such perfect electron confinement is also unrealistic given the topology of the magnetic fields of the small magnets in the experiment, where the field lines also end at surfaces, leading to a loss of electrons. This is mimicked in the code by restricting the magnetic field only to the vicinity of the slits so that electron losses to the surface can occur outside the slit regions.\\



\begin{figure}
    \centering
    \includegraphics[width=0.6\textwidth]{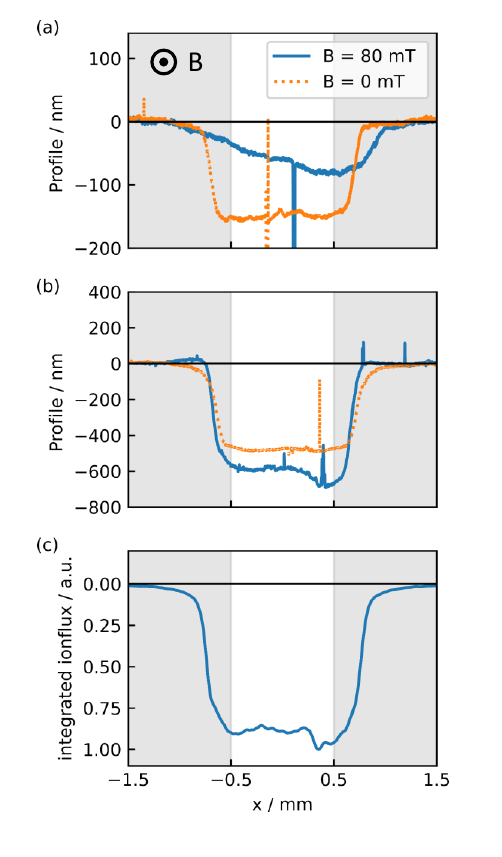}
    \caption{3D etching profiles of C:H in an argon-oxygen plasma with -100\,V self-bias (a) and of silicon in a C:F plasma with -150\,V self-bias (b) as well as the integrated ion flux from the simulation at 100\,V applied voltage (c).} 
    \label{fig:chetching}
\end{figure}

In the first experiment, the etching of a hydrocarbon film (C:H) in an argon-oxygen plasma at an RF self-bias of -100\,V is analyzed. Two experiments with (blue solid line) and without (orange dashed line) a magnetic field are compared and shown as an etch profile across the slit in Fig. \ref{fig:chetching}a. The grey area indicates the position of the mask above the wafer, and the direction of the magnetic field is out of the plane, as indicated. 

With a magnetic field, one can observe that the etch rate on the right-hand side is much larger than on the left-hand side (blue solid line in Fig. \ref{fig:chetching}a). This seemingly contradicts what one would expect, given that the Lorentz force should deflect the incident ion towards the left-hand side. Additionally, the $\vec{E}\times\vec{B}$ drift of the electrons, for an electric field perpendicular to the substrate, should also lead to a deflection of the plasma to the left-hand side.   

In a second experiment, the asymmetry of Si etching in a CF$_4$ plasma at an RF self-bias of -150\,V is analyzed. Two experiments with and without the magnetic field are compared and shown as an etch profile across the slit in Fig. \ref{fig:chetching}b. One can see an asymmetry in the etch profile similar to the experiments on the C:H film etching.
Additionally, a local maximum in the center of the trench can be observed. XPS analysis (not shown) revealed the deposition of a C:F polymer in the center of the trench.

\begin{figure*}
    \centering
    \includegraphics[width=\columnwidth]{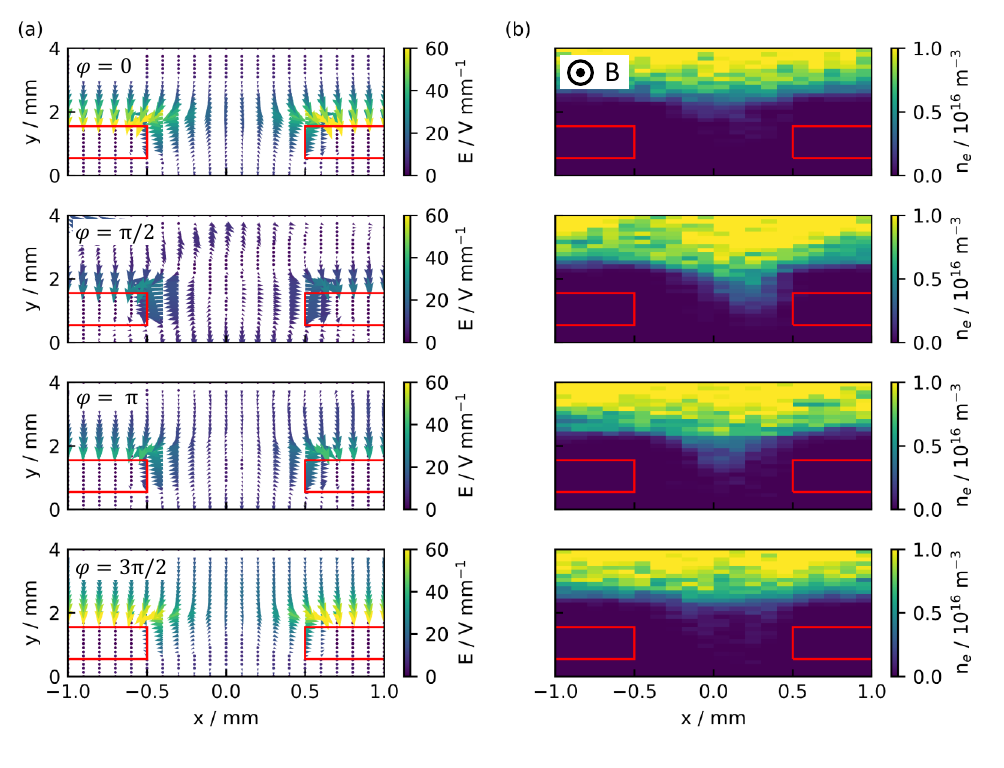}
    \caption{The space-resolved electric field (a) and the electron density (b) from the PIC simulation for an RF amplitude of 100~V at different phases of the RF cycle, as indicated. The red rectangles represent the mask.}
    \label{fig:picrfcycle}
\end{figure*}

The integrated ion flux below the mask structure from the PIC/MCC simulation for an applied voltage of 100\,V is shown in  Fig. \ref{fig:chetching}c. The integrated flux is also higher on the right-hand side of the slit. This is in agreement with the experiment. The peaks at the edges of the profile are caused by the glancing angle reflection of ions on the inside walls of the mask. However, any quantitative  comparison of the profiles from simulation and experiment is limited, 
since the simulation investigates only the ion flux and does not yet include the chemical sputtering process to obtain an etch rate. 
Furthermore, the space- and time-resolved electric field and electron dynamics in the mask region are investigated. 
Fig. \ref{fig:picrfcycle} shows the space-resolved electric field (a) and electron densities (b) at different phases relative to the RF cycle. 
The mask structure modifies the electric field, resulting in the electric field vectors inside the slit pointing towards the interior of the mask structure parallel to the substrate below.
The plasma penetrates the slit at the different phases of the RF cycle, predominantly at $\phi=\pi/2$ when the sheath collapses, and the sheath voltage becomes minimal. At $\phi=3/2 \pi $, the plasma is expelled from the slit due to maximum sheath expansion. It is interesting to regard the phase of maximum plasma penetration at $\phi=\pi/2$. Here, one can see that the plasma penetrates asymmetrically into the slit opening. One could also note that the plasma density shows some structure above the mask. By inspecting the time development of the density maps, one can observe the propagation of waves along the interface. Based on the geometry of the magnetic field and density gradient, these waves are presumably gradient drift waves.

The experiment revealed asymmetry in the etching process due to a magnetic field and mask in front of a wafer. Based on the comparison between the experiment and modeling, two processes overlap, as illustrated in Fig. \ref{fig:model}:

\begin{figure}
    \centering
    \includegraphics[width=0.6\textwidth]{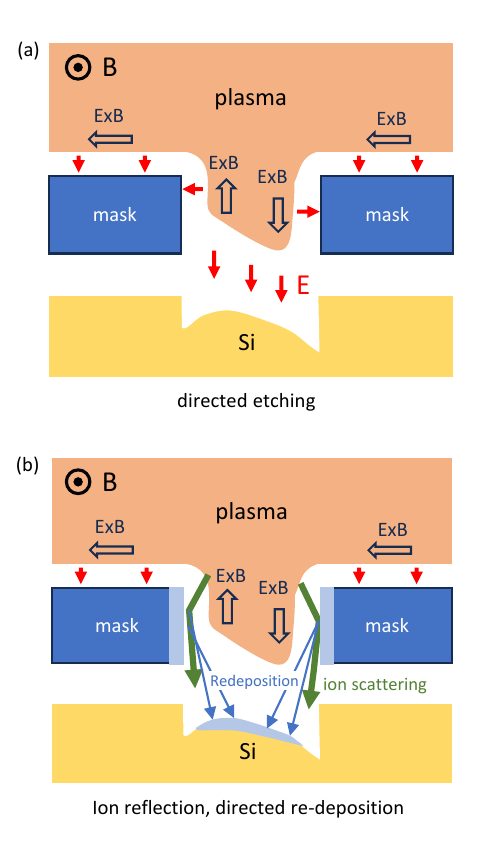}
    \caption{Mechanisms to create a 3D deposition or etching profile: (a) varying penetration of the plasma into the mask structure depending on the  $\vec{E} \times \vec{B}$ drifts, (b) ion focusing and re-deposition from the mask.}
    \label{fig:model}
\end{figure}

\begin{itemize}
\item\textbf{Plasma insertion:} The plasma penetrates the metal slit structure, and an electric field develops between the plasma and the mask sidewalls. This electric field in the plasma sheath is \textit{parallel} to the wafer surface and causes an $\vec{E} \times \vec{B}$ drift on the electrons either towards the wafer or away from the wafer. As a result, the plasma penetrates further into the slit of the mask on one side of the slit compared to the other (Fig. \ref{fig:model}a).

This plasma insertion should scale with the sheath thickness in relation to the slit width of the mask. If we take the simple Child-Langmuir sheath model and an electron temperature of 3\,eV and a plasma density of 10$^{17}$\,m$^3$, we obtain a sheath thickness of 0.6\,mm at a bias voltage of -150\,V. Then the plasma may not be able to penetrate the slit structure completely. However, the RF sheath is modulated, and within an RF cycle, the plasma partially enters the slit structure when the sheath collapses. 

\item\textbf{Ion scattering, re-deposition:} The etch profile is further modified by glancing angle scattering of ions at the inside wall of the metal mask. In addition, the re-deposition of any deposits from the inside walls of the mask causes re-deposition on the substrate surface (Fig. \ref{fig:model}b). A typical example is C:F polymer re-deposition during CF$_4$ etching of silicon.
\end{itemize}

This effect can be controlled by changing the magnetic field or any electric field by biasing these masks, but also by changing the plasma properties, which affects the relation between the sheath thickness and mask feature sizes since the guiding length scale is the plasma sheath thickness which ranges for typical plasma densities of at least 10s of microns or larger. The magnetic field's impact is crucial for the $\vec{E}\times\vec{B}$ drift but also modifies the sheath thickness. Since the magnetic field directs parallel to the wafer surface, the confinement of electrons is improved so that the effective bias voltage is smaller than in the non-magnetized case.\\


In conclusion, it can be stated that adding a magnetic field to an etch mask in front of a wafer induces an asymmetry in the incident ion flux onto the wafer. The effect is caused by asymmetric penetration of the plasma into the mask structure, due to the $\vec{E}\times \vec{B}$ drift, which increases the penetration of the plasma on one side of the mask and decreases it on the other side. This effect depends sensitively on the relation between the mask's length scale and the plasma sheath's length scale to allow plasma penetration. The exploitation as a tool to also asymmetrically etch depends on the etch chemistry of chemical sputtering, which was very strong in the case of C:H etching, leading to very asymmetric etch profiles but only small in the case of silicon etching. \\
In the future, it will be necessary to further explore this effect experimentally to optimize the etch asymmetry and complementary develop the PIC/MCC model by including the chemistry directly and a realistic surface model for the different materials which includes the reflection of particles and the emission of secondary electrons.

\begin{acknowledgments}
The authors thank Martin Hoffmann (Microsystem Technology, Ruhr University Bochum, Germany) and Stefan Sinzinger (Technical Optics, Technical University Ilmenau, Germany) for helpful discussions.
\end{acknowledgments}

\section*{Conflict of interest statement}

The authors have no conflicts to disclose.

\section*{Data Availability Statement}

Data are available on request from the authors.

\bibliography{Proposal3detching}

\end{document}